\documentclass{llncs}
\usepackage{makeidx}  
\usepackage{graphicx} 
\begin{document}
%
%
\pagestyle{headings}  
%
\mainmatter             
\title{SHARE: A Web Service Based Framework for Distributed Querying and Reasoning on the Semantic Web}
\titlerunning{SHARE: Distributing Querying and Reasoning Framework}  
%
\author{Ben P Vandervalk \and E Luke McCarthy \and Mark D Wilkinson}
\institute{The Providence Heart + Lung Research Institute at St. Paul's Hospital,\\
University of British Columbia, Department of Medical Genetics, \\
Vancouver, BC, Canada \\
\email{markw@illuminae.com} }

\maketitle              

\begin{abstract}
Here we describe the SHARE system, a web service based framework for distributed querying and reasoning on the semantic web.  The main innovations of SHARE are: (1) the extension of a SPARQL query engine to perform on-demand data retrieval from web services, and (2) the extension of an OWL reasoner to test property restrictions by means of web service invocations.  In addition to enabling queries across distributed datasets, the system allows for a target dataset that is significantly larger than is possible under current, centralized approaches.  Although the architecture is equally applicable to all types of data, the SHARE system targets bioinformatics, due to the large number of interoperable web services that are already available in this area.  SHARE is built entirely on semantic web standards, and is the successor of the BioMOBY project.
\end{abstract}

\section*{Introduction}

The vision of the semantic web is to build a massive network of distributed, interconnected, machine-readable data \cite{1}\cite{2}. The goal is not only for software programs to be able to access and query the data itself, but also to make automated inferences based on the meaning that is encoded therein.  The core components of the semantic web have now been established by the W3C: we have RDF \cite{3}, a language for describing data; OWL \cite{4}, a language for defining ontologies; and SPARQL \cite{5}, a language for querying RDF.  In addition, several OWL reasoners \cite{6}\cite{7}\cite{8} have been implemented which are capable of classifying data when given an ontology and a set of instance data. 

Unfortunately, crucial infrastructure for querying and reasoning across distributed datasets is still missing.  Current SPARQL implementations handle remote data sets by downloading them to the site of the query engine in their entirety \cite{9}, and reasoners are likewise dependent on a single, centralized dataset.  In the realm of bioinformatics, a distributed framework for querying and reasoning would be particularly valuable.  There are now more than a thousand biological databases on the web \cite{10}, containing distinct but fundamentally interrelated information about DNA sequences, protein structures, networks of metabolic reactions, chemical properties of molecules, and so on.  The need for a simple and effective means of integrating these databases is evidenced by the numerous publications \cite{11}--\cite{14}, data warehouses \cite{15}--\cite{18}, and software systems \cite{19}--\cite{25} that have been inspired by the problem.  One such system is BioMOBY; the SHARE project described here upgrades and extends BioMOBY, creating a general purpose architecture for querying and reasoning over the semantic web.

\section*{Past Work: BioMoby}

BioMoby\footnote{Moby is not an acronym, it's just a name.  The name comes from the conference where the idea was conceived: MOBY-DIC (Model Organism Bring Your Own Database Interface Conference).}  is a simple framework for defining and discovering interoperable web services.  Although Moby is a generic solution which can be applied to any type of service, bioinformatics is the area in which it is currently being used.  Under Moby, services communicate according to a shared messaging format, and all inputs and outputs of services are specified in terms of a centralized Moby datatype ontology.  This ontology defines both syntax and semantics for a large number bioinformatics datatypes such as DNA sequences, Gene Ontology \cite{26} terms, Single Nucleotide Polymorphisms (SNPs), and so on.  For example, the object for representing a protein sequence is called \textbf{AminoAcidSequence} and has two member values:  an integer for storing the length of the sequence, and a string for storing the sequence itself. Each datatype specifies its own serialization into XML, and new datatypes may be introduced by any user of the system. The precise specification of datatypes allows services to be easily chained into \emph{workflows}, in which the output of one service becomes the input of the next.

In addition to a datatype ontology, Moby also maintains a large working registry of services.  The registry now holds approximately 1500 web services which perform a wide variety of tasks such as database retrieval, alignment of sequences, identification of protein domains, prediction of subcellular localization, etc.  The most important feature of the Moby registry is the ability to query for services by input or output datatype.  This enables the stepwise, interactive construction of workflows which perform complex analyses.  Moby workflows may be constructed in a GUI environment such as Taverna \cite{27}, or executed immediately as they are traversed, by means of a client such as GBrowse Moby \cite{28}.

\begin{figure}
\centering
\includegraphics[width=5.5in,natwidth=3600,natheight=2901]{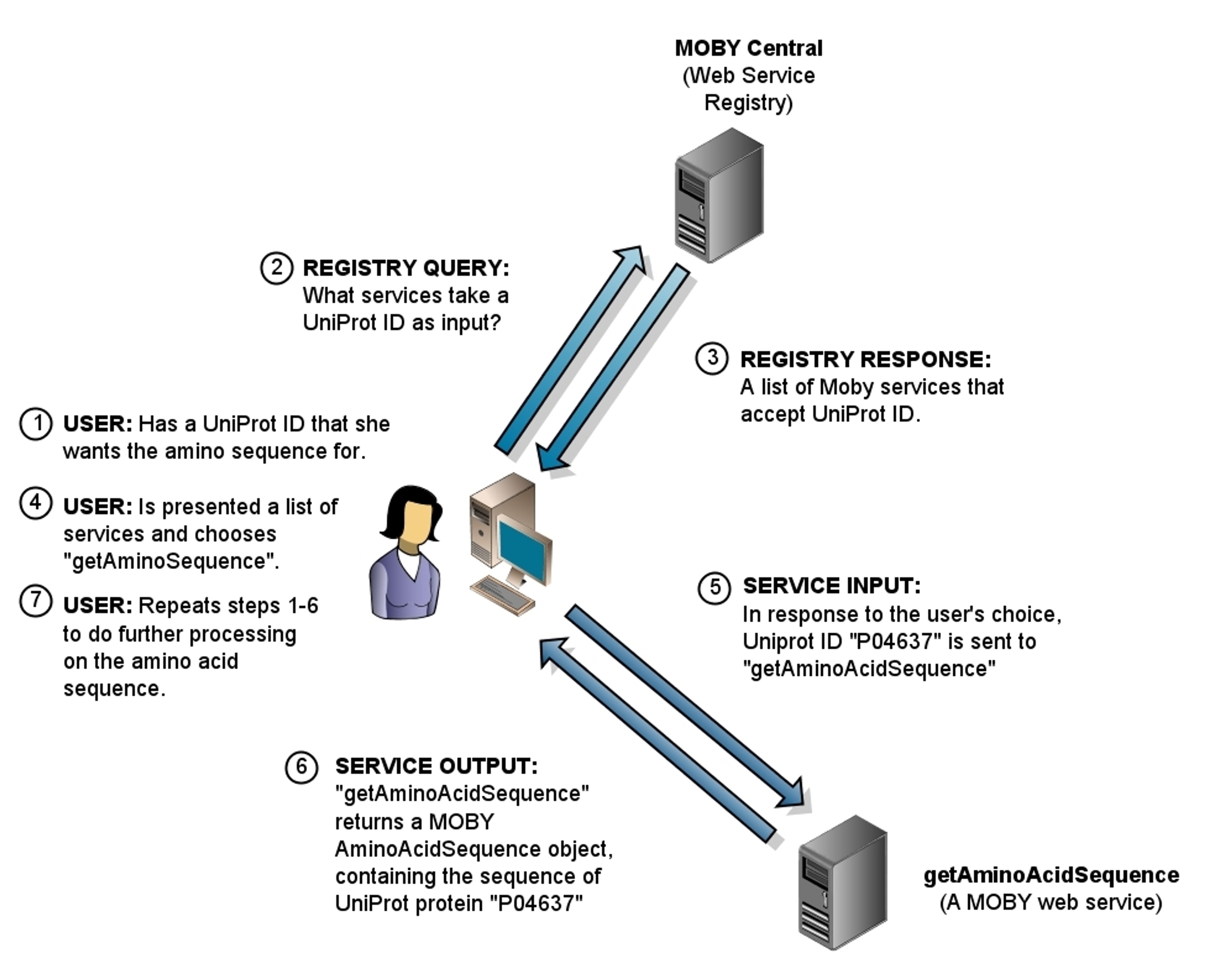}
\caption{Typical usage of the BioMOBY framework.  (1) The user begins with data that matches a certain Moby datatype.  Usually this data is a bare identifier, which corresponds to the default Moby datatype \textbf{Object}.  (2) The user queries the registry for services that consume her identifier as input.  (3) The registry returns a list of such services.  (4) The user chooses a service from the list, based on the desired type of analysis. (5) The user's data is sent to the chosen service, in this case \textbf{getAminoAcidSequence}, and the service is executed.  (6) The service returns its output, in this case a data object of type \textbf{AminoAcidSequence}.  (7) The user repeats steps 1-6, until the desired analysis of the data is complete.  The reader may try steps 1-6 using the GBrowse Moby client at http://moby.ucalgary.ca/gbrowse\_moby.}
\label{fig:mobyframework}
\end{figure}

The Moby architecture is depicted in Figure \ref{fig:mobyframework}.

\section*{Recent Work: SPARQL Queries Resolved By Web Services}

One of the main limitations of BioMoby is its reliance on a custom XML format, making it difficult for Moby services to be used within other frameworks.  Unfortunately, the invention of an extensible data syntax was necessary as BioMOBY predates the advent of RDF.  SHARE is a major revision of the MOBY framework which corrects this shortcoming and establishes a completely generic, open framework based on semantic web standards.  At the same time, SHARE introduces higher-level querying and reasoning functionality.

\begin{figure}
\centering
\includegraphics[width=5.5in, natwidth=3300,natheight=1523]{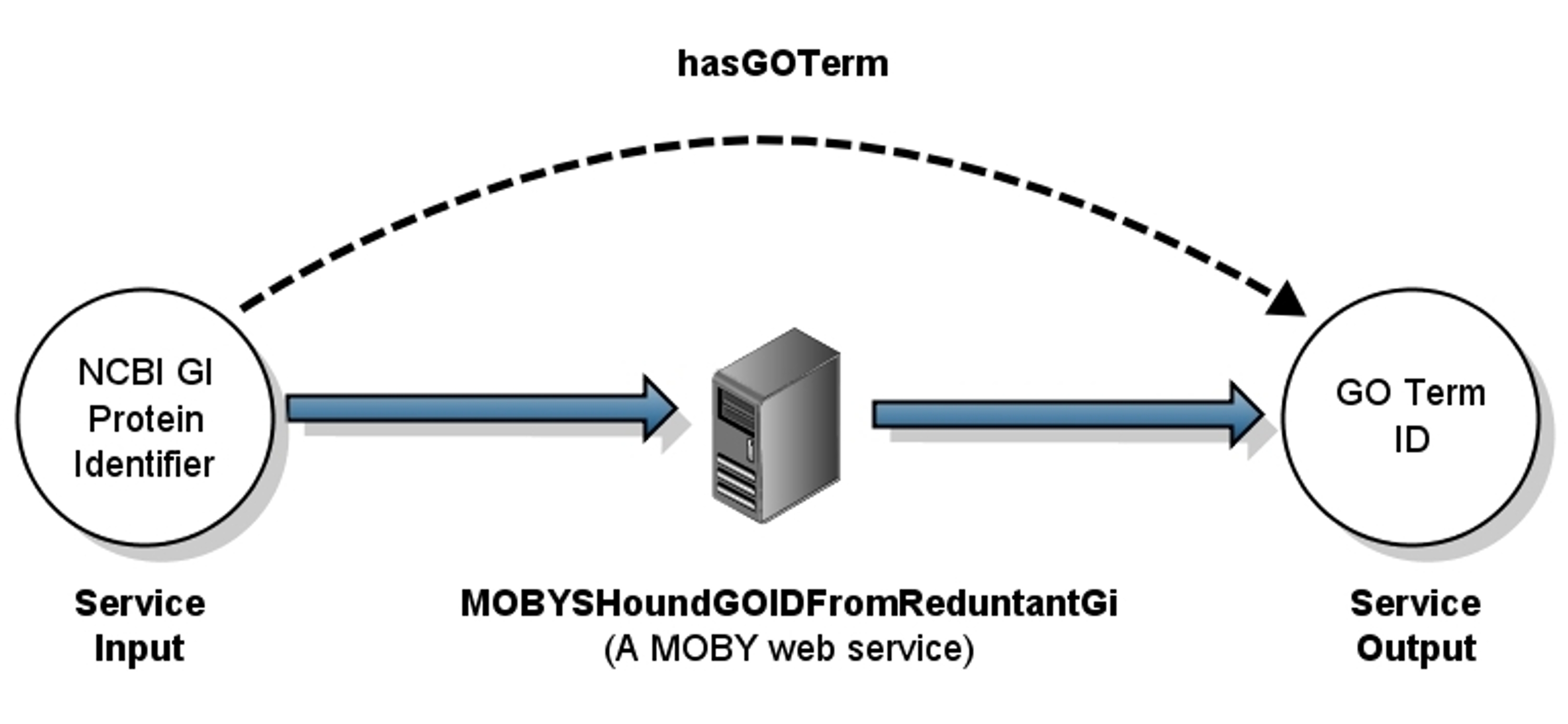}
\caption{The key observation behind the SHARE framework: a web service invocation generates an implicit RDF triple.  The subject of this triple is the input, the object is the output, and the predicate is the relationship established between the input and output, as determined by the behaviour of the service.   In this case, the service consumes a GI (Genbank Identifier) for a protein, and returns one or more GO terms which annotate the protein.  The implicit relationship is \textbf{hasGOTerm}.}
\label{fig:servicetriple}
\end{figure}

The SHARE system is based on the following key observation: whenever a web service computes a result, it is in effect generating an RDF triple.  The subject of this triple is the input, the object is the output, and the predicate is the relationship that is established between the input and the output by the service call.  In other words, the predicate is defined by the behaviour of the service.   For example, a service that retrieves a list of GO (Gene Ontology) annotations for a protein generates triples of the form ``$<$protein ID$>$ hasGOTerm $<$GO term ID$>$'', as shown in Figure \ref{fig:servicetriple}.  It is logical then, to annotate the service itself with the predicate \textbf{hasGOTerm}.\footnote{More accurately, a predicate annotation connects one input and one output of a service.  A Moby service may have arbitrarily many inputs and outputs, with differing datatypes.}

The system provides a specialized SPARQL engine which utilizes these predicate annotations to retrieve data ``on demand'' from web services.  The syntax of a SHARE query is identical to that of a standard SPARQL query, with the only difference being the resolution behaviour.  A query is resolved by: (1) identifying any predicates that can be matched to services, (2) retrieving data from these services, and (3) allowing the query to be resolved as usual on the local triple store. Figure \ref{fig:querypark} shows an example query which asks: ``What transcription factors have been implicated in Parkinson's Disease?''.

\begin{figure}

{\tt
SELECT ?transcriptionFactor\\
WHERE\\
\{\\
    ?transcriptionFactor SHARE:hasGOTerm GO:0006351 .\\
    ?transcriptionFactor SHARE:associatedWithDisease OMIM:168600 .\\
\}
}
\caption{A hypothetical SHARE query, which finds transcription factors implicated in Parkinson's Disease.  Supposing both \textbf{hasGOTerm} and \textbf{associatedWithDisease} have been assigned to services, the proteins with the specified predicate values can be retrieved dynamically via web service invocations.}
\label{fig:querypark}
\end{figure}

SHARE depends on access to a large central registry of services which are annotated with appropriate predicates.  This registry is provided by the existing BioMoby framework and community.  Services participating in the SHARE system are required to follow two simple rules: (1) All inputs and outputs of services must be RDF documents, and (2) All inputs and outputs must be specified in terms of OWL classes.  A ``seed'' ontology of OWL classes will be provided based on existing BioMoby datatypes, but the system will be completely open to expansion; service providers may specify their interfaces in terms of any OWL classes they choose.  The use of OWL to specify interfaces, rather than WSDL \cite{29}, will enable description of both the syntax \emph{and the meaning} of service arguments, thus allowing for a community of truly interoperable services.  In addition, service providers will be encouraged to supply predicate annotations for their services.  However, as it does no harm to assign multiple predicates to the same service, any users of the system will be able to assign predicates as well.

An early prototype of SHARE, with example queries, is accessible at \\ http://cardioshare.icapture.ubc.ca/cardioSHARE/query.

The system represents a valuable enhancement to standard query systems, as it offers a straightforward mechanism for querying across any number of data sources.  In effect, the target of a SHARE query is an enormous \emph{virtual graph}, consisting of all triples that can generated by the complete set of participating services.\footnote{This includes the full set of $\sim$1500 BioMoby services already in the system.}  Beyond providing a large, integrated dataset, the system has several additional advantages.  As a web service based framework, participating services need not be simple retrieval mechanisms for data; they are capable of performing any calculation that can be accomplished by software.  SHARE is therefore not only a framework for integrating databases, but also a framework for integrating analytical programs.  A further advantage of the system is that new services may be added by anyone, and the responsibility for maintaining these services is distributed to their creators.

Intuition might suggest that SHARE queries, because they must retrieve data from many remote sources, are vastly slower than equivalent queries on a data warehouse.  This is not necessarily the case.  For example, one important optimization trick for speeding up query resolution is the use of \emph{inverse services}.  Considering the example query in Figure \ref{fig:querypark}, the system might naively find proteins that are associated with Parkinsons (OMIM:168600) by feeding every known protein into a web service that returns OMIM codes.  However, it is equally possible that there is a service which accepts OMIM codes as input and return associated proteins.\footnote{In fact, there is such a service in the BioMoby registry, and it is called \textbf{MOBYSHoundGiFromOMIM}.}  In the latter case, the question can be answered with a single service invocation.

\section*{Current Work: DL Reasoning Resolved By Web Services}

In a similar fashion, the SHARE framework will extend an OWL reasoner to use predicate annotations on services.  When determining instances of a class, the reasoner will have the ability to test property restrictions by means of web service invocations.  For example, we could define an OWL class called \textbf{ParkinsonTranscriptionFactor} with the restrictions (hasGOTerm hasValue GO:0006351) and (associatedWithDisease hasValue OMIM:168600).  We could then answer the question posed in the previous section, by finding instances of this class.  This is completely equivalent to the SPARQL query posed in Figure \ref{fig:querypark}.

The SPARQL and DL reasoning aspects of SHARE will be tied together by allowing an OWL class to be referenced within a SPARQL query.  This facility will allow users to formulate complex queries in simple, abstract language. For instance, the original query in Figure \ref{fig:querypark} could be extended to find transcription factors which are both implicated in Parkinson's disease and also have experimentally solved 3D structures (Figure \ref{fig:querypark2}).

\begin{figure}
{\tt
SELECT ?transcriptionFactor\\
WHERE\\
\{\\
    ?transcriptionFactor rdf:type SHARE:ParkinsonTranscriptionFactor .\\
    ?transcriptionFactor SHARE:hasSolved3DStructure ?structure .\\
\}
}
\caption{A hypothetical SHARE query, which finds transcription factors that are both implicated in Parkinson's Disease and have at least one experimentally solved 3D structure.  The \textbf{rdf:type} triple tells the query engine to match \textbf{?transcriptionFactor} to instances of the OWL class \textbf{ParkinsonTranscriptionFactor}.  The system retrieves these instances by invoking web services corresponding to the predicates (\textbf{hasGOTerm} and \textbf{associatedWithDisease}) that have been used to define the class.  Each of the instances is then sent to one or more web services that have been annotated with \textbf{hasSolved3DStructure}, in order to retrieve any solved 3D structures that are available.}
\label{fig:querypark2}
\end{figure}

It is reasonable to ask what purpose the reasoner extension serves if classification is exactly like querying, but with the additional restrictions imposed by OWL-DL.  The advantage of the reasoner approach can be seen if one imagines defining classes in terms of other classes. If instead of being defined by specific URI values for properties, \textbf{ParkinsonTranscriptionFactor} was defined by the intersection of  \textbf{ParkinsonAssociatedProtein} and \textbf{TranscriptionFactor} classes, each having a long list of property restrictions, the equivalent SPARQL query would likely be quite complex.  The use of OWL classes provides modularity, reusability, and simplicity when formulating queries.

In addition to enabling reasoning across distributed data sources, the SHARE reasoner will enable classification over large-scale datasets without the need to make changes to existing reasoning algorithms.  This is possible for the same reason that large-scale SPARQL queries are possible; the use of inverse services (as explained above) filters out large amounts of irrelevant data that would otherwise have to be processed by the query engine or reasoner.

From a bioinformatics perspective, one of the most interesting applications of the SHARE reasoner will be its ability to automatically ``lift'' raw data into an ontology.  If a user wants to gather a complete list of instances for each class in an ontology, all they will have to do is assign the properties of the ontology to available web services, and then run the reasoner.  This is interesting because the majority of data annotation in bioinformatics is still done manually with controlled vocabularies such as the Gene Ontology.

The first application of SHARE will be in the analysis of clinical data relating to heart disease.  This will entail the development of a SHARE ontology to encode expert knowledge about cardiovascular disease.  The research environment provided by this ontology, together with the SHARE framework, will be called CardioSHARE.

\section*{Conclusion}

Currently there are no widely accepted systems for querying or reasoning across distributed data sources.  The SHARE framework provides these capabilities, by means of simple extensions to existing query engines and reasoners.   At the same time, SHARE allows these tools to operate on vastly larger datasets than would otherwise be possible.  The price that must be paid for achieving these improvements is typical of data integration projects in general.  First, the system must gain widespread community support in order to have any true value for its users. Fortunately, we already have access to a large community of service providers and users, through the legacy of the BioMoby system.  Secondly, service providers must play by a shared set of rules.  In the case of SHARE, the rules are simple: the inputs and outputs of services must be RDF documents that are described by OWL classes.

\section*{Acknowledgements}

The development of SHARE and CardioSHARE is made possible by the support of the Heart and Stroke Foundation of British Columbia and Yukon. MDW is funded for CardioSHARE through an operating grant from the Canadian Institutes of Health Research. BioMOBY was developed under support from the Genome Canada/Genome Alberta bioinformatics Platform. Hardware for both projects has been provided by Sun Microsystems and IBM. Core laboratory funding is provided by the Natural Sciences and Engineering Research Council of Canada (NSERC).

%
%

\end{document}